\documentclass[pdftex,final]{ifacmtg}
\usepackage{graphicx,latexsym,blackboard}

\def\H{{\cal H}}
\def\L{{\cal L}}

\def\boldsymbol#1{\mathbf{#1}}
\def\op#1{\hat{#1}}
\def\ket#1{| #1 \rangle}

\def\bra#1{\langle #1 |}

\def\trace{\mathop{\rm Tr}\nolimits}

\begin{document}
\begin{frontmatter}
\title{Controllability of quantum systems}
\author{Sonia G.\ Schirmer}
\vspace{-2ex}
\address{Dept of Applied Mathematics + Theoretical Physics and Dept of Engineering,
         University of Cambridge, \\ Cambridge, CB2 1PZ, United Kingdom}
\author{Ivan C.~H.\ Pullen and Allan I.\ Solomon}
\vspace{-2ex}
\address{Quantum Processes Group, The Open University, \\ 
         Milton Keynes, MK7 6AA, United Kingdom} 
\vspace{-2ex}
\begin{abstract}
An overview and synthesis of results and criteria for open-loop controllability of 
Hamiltonian quantum systems obtained using dynamical Lie group and algebra techniques
is presented.  Negative results for open-loop controllability of dissipative systems
are discussed, and the superiority of closed-loop (feedback) control for quantum 
systems is established. 
\end{abstract}
\end{frontmatter}

\section{Introduction}

Controlling the dynamics of quantum systems has been a long-standing goal of quantum
physicists and chemists, which has received renewed attention recently driven by the
desire to build quantum information processing devices.  A central problem one faces
when attempting to control quantum systems, is the question to which extent it is 
possible to control the dynamics of the system such as to achieve a certain aim or 
control objective.

In the following we first provide a synthesis of recent results on controllability of 
finite-dimensional closed quantum systems subject to open-loop coherent control, i.e.,
coherent control without measurements and feedback.  The results show that for these
systems open-loop control is sufficient for controllability in most cases.  For open
systems, i.e., systems subject to dissipative effects due to uncontrollable interactions
with their environment, however, it has been shown that open-loop control is generally
not sufficient for controllability, even in finite dimension.  We discuss the negative
controllability results in this case.  Finally, we present a brief overview of recent
results using closed-loop control of quantum systems, i.e., control involving 
measurements and feedback, which indicate its superiority over open-loop control in 
various settings. 

\section{Mathematical Model}

We consider the problem of control of quantum systems whose state space is a Hilbert
space $\H$ of dimension $N$.  The state of such a system can be described by a density
operator $\op{\rho}$.
\begin{defn}
A \emph{density operator} is a positive operator on $\H$ with unit trace.  If it has
rank one it is said to represent a \emph{pure state}, otherwise it represents a 
\emph{mixed state}.
\end{defn}
A density operator representing a pure state is simply a projector onto a one-dimensional
subspace of the Hilbert space $\H$, i.e., using Dirac notation we can write $\op{\rho}=
\ket{\Psi}\bra{\Psi}$, where $\ket{\Psi}$ is an element in $\H$ called a wavefunction.
Thus, the special case of a pure state can also represented by a wavefunction $\ket{\psi}
\in\H$.

In general the operator $\op{\rho}$ satisfies the quantum Liouville equation
\begin{equation}
  i\hbar \frac{d}{dt} \op{\rho}(t) 
  = \left[\op{H}[\vec{f}(t)],\op{\rho}(t)\right] + i\hbar\L_D[\op{\rho}(t)],
\end{equation}
where the commutator determines the Hamiltonian part of the evolution, and the second
term represents the non-Hamiltonian part due to dissipative effects, which may include
certain measurments that weakly perturb the system.  The operator $\op{H}[\vec{f}(t)]$
is the total Hamiltonian of the system subject to the control fields $\vec{f}(t)$. For
the purpose of controllability studies the control Hamiltonian is usually assumed to 
be control-linear, i.e.,
\begin{equation} \label{eq:H}
  \op{H}[\vec{f}(t)] = \op{H}_0 + \sum_{m=1}^M f_m(t)\op{H}_m,
\end{equation}
where $\op{H}_0$ is the internal system Hamiltonian and $\op{H}_m$, $m>0$, are the
interaction terms.  This assumption is generally reasonable as long as the control
fields are sufficiently weak compared, e.g., to relevant intra-atomic or molecular 
forces.

In the absence of dissipative effects, i.e., when $\L_D=0$, the evolution of the system
is Hamiltonian and the density operator satisfies the dynamical evolution equation
\begin{equation}
   \op{\rho}(t) = \op{U}(t,t_0) \op{\rho}_0 \op{U}(t,t_0)^\dagger,
\end{equation}
where $\op{U}(t,t_0)$ is the Hilbert space evolution operator that satisfies the 
Schrodinger equation
\begin{equation}
  i\hbar \frac{d}{dt} \op{U}(t,t_0) = \op{H}[\vec{f}(t)] \op{U}(t,t_0),
\end{equation}
and is hence restricted to the unitary group $U(N)$, $N$ being the dimension of the
Hilbert space $\H$. 

The constraint of unitary evolution for Hamiltonian systems imposes restrictions on 
the quantum states that can be reached from a given initial state.  Since the former
limitations are independent of the control fields and their type of interaction with
the system, we shall refer to them as \emph{kinematical constraints}, to distinguish 
them from \emph{dynamical constraints} arising from constraints on the control fields
or the interaction with those fields.
Precisely, given an initial state $\op{\rho}_0$ and a target state $\op{\rho}_1$, 
the target state is kinematically admissible if and only if there exists a unitary
operator $\op{U}$ such that $\op{\rho}_1 = \op{U} \op{\rho}_0 \op{U}^\dagger$. This
equation defines an equivalence relation that partitions the set of density operators
into kinematical equivalence classes.  Two density matrices $\op{\rho}_1$, $\op{\rho}_2$
are \emph{kinematically equivalent} if they are unitarily equivalent, i.e., if 
$\op{\rho}_2=\op{U}\op{\rho}_1\op{U}^\dagger$ for some unitary operator $\op{U}$.   
It follows immediately that two density matrices are kinematically equivalent if and
only if they have the same eigenvalues.  Hence, the set of pure states forms one such
equivalence class of states.
 
\section{Degrees of Controllability for Closed Systems}

Closed quantum systems, i.e., systems that are decoupled from their environment except
for (controlled) interactions with coherent control fields, exhibit Hamiltonian dynamics
if they are subjected to open-loop coherent control.  Various notions of controllability
exist for such systems.  Depending on the application, one may be interested in complete 
controllability, observable or operator controllability, and mixed-state or pure-state
controllability [\cite{qph0106128,JPA35p4125}].
\begin{defn}
A quantum system subject to Hamiltonian dynamics is \emph{completely controllable} 
if any unitary evolution is dynamically realizable, i.e., if for any unitary operator
$\op{U}$, there exists an admissible control-trajectory pair $(\vec{f}(t),\op{U}
(t,t_0))$ defined for $t_0\le t\le t_F$ (for some $t_F<\infty$) such that $\op{U}=\op{U}(t_F,t_0)$.
\end{defn}
\begin{defn}
A quantum system subject to Hamiltonian dynamics is \emph{density matrix controllable}
if given any pair of kinematically equivalent density matrices $\op{\rho}_0$ and 
$\op{\rho}_1$, there exists an admissible control-trajectory pair $(\vec{f}(t),
\op{U}(t,t_0))$ defined for $t_0\le t\le t_F$ (for some $t_F<\infty$) such that 
$\op{\rho}_1=\op{U}(t_F,t_0) \op{\rho}_0 \op{U}(t_F,t_0)^\dagger$. 
\end{defn}
\begin{defn}
A quantum system subject to Hamiltonian dynamics is \emph{pure-state controllable} 
if for any two pure states given by the (normalized) wavefunctions $\ket{\Psi_0}$, 
$\ket{\Psi_1}$, there exists an admissible control-trajectory pair $(\vec{f}(t),
\op{U}(t,t_0))$ defined for $t_0\le t\le t_F$ (for some $t_F<\infty$) such that 
$\ket{\Psi_1}=\op{U}(t_F,t_0)\ket{\Psi_0}$.
\end{defn}
\begin{defn}
A quantum system subject to Hamiltonian dynamics is \emph{observable controllable} 
if for any observable $\op{A}$ and initial state $\op{\rho}_0$ of the system, there
exists an admissible control-trajectory pair $(\vec{f}(t),\op{U}(t,t_0))$ defined 
for $t_0\le t\le t_F$ (for some $t_F<\infty$) such that the ensemble average 
$\trace[\op{\rho}(t)\op{A}]$ of $\op{A}$ assumes any kinematically admissible 
value [\cite{PRA58p2684}].
\end{defn}

\section{Lie-algebraic criteria for controllability}

The notions of controllability for closed quantum systems subject to open-loop coherent
control defined in the previous section can be related to the dynamical Lie group and
Lie algebra of the control system.  
\begin{defn}
The \emph{dynamical Lie algebra} of a quantum system with Hamiltonian (\ref{eq:H}) is 
the Lie algebra $\L$ generated by the skew-Hermitian operators $i\op{H}_m$ by taking
all linear combinations and iterated commutators.
\end{defn}
The associated dynamical Lie group $G$ is formed by the elements $\exp(\op{x})$ where 
$\op{x}$ is an element of $\L$.  Knowledge of the dynamical Lie group of the system
is fundamental to understanding the dynamics of the system and especially limitations
on its control since the dynamical evolution operator $\op{U}(t,t_0)$ of the system is 
constrained to this Lie group for all times, independent of the control fields applied,
and the set of reachable states and dynamically realizable bounds on the expectation
values of observables depend on this dynamical Lie group.

For Hamiltonian quantum systems it is obvious from the definition that only a system 
with dynamical Lie group $U(N)$, where $N$ is the dimension of the Hilbert space $\H$, 
can be completely controllable.  Density matrix controllability requires that the 
dynamical Lie group act transitively on \emph{all} equivalence classes of density 
matrices, while pure-state controllability requires only transitive action of the 
dynamical Lie group on the equivalence class of pure states.  Since the latter can 
be represented by normalized wavefunctions or unit vectors in $\CC^N$, pure-state 
controllability requires only transitive action of the dynamical Lie group on the 
unit sphere in $\CC^N$.  

It can furthermore be shown that density matrix controllability is both necessary 
and sufficient for observable controllability as defined above [\cite{JPA35p4125}].  
Based on these observations, the degree of controllability of a Hamiltonian quantum 
system with control-linear Hamiltonian (\ref{eq:H}) can be characterized in terms 
of its dynamical Lie algebra:
\begin{thm} \label{thm:a} [\cite{qph0106128, JPA35p4125}]
A Hamiltonian quantum system with control-linear Hamiltonian is
\begin{itemize}
\item completely controllable if and only if $\L\simeq u(N)$.
\item density matrix controllable if and only if $\L\simeq su(N)$ or $\L\simeq u(N)$.
\item pure-state controllable if and only if $\L$ is isomorphic to either
      $u(N)$, $su(N)$, or (when $N=2\ell$) $sp(\ell)$, $sp(\ell)\oplus u(1)$.
\end{itemize} \vspace{-2ex}
\end{thm}
Complete controllability thus implies density matrix and observable controllability 
and the latter guarantees pure-state controllability.  According to our definition
complete controllability is strictly speaking a slightly stronger requirement than 
density matrix or observable controllability but the difference is essentially only
a phase factor, which is not important for most applications.  
\begin{defn}
For simplicity we shall therefore call a system \emph{controllable} if it is density 
matrix and observable controllable.
\end{defn}
An interesting consequence of the previous result is that pure-state controllability is 
equivalent to mixed-state controllability if the dimension of $\H$ is odd, but it is a 
weaker requirement if the dimension of $\H$ is even, i.e., there exist even-dimensional
quantum systems that are pure-state controllable but not mixed-state controllable.

For a system that is not controllable there exist initial states for which the set of
dynamically accessible target states does not comprise the entire kinematical equivalence
class, i.e., at least some of the kinematical equivalence classes are partitioned into 
disjoint subsets of dynamically equivalent states.  If the system fails to be pure-state 
controllable then there exists such a partition for the equivalence class of pure states
[\cite{IEEE40CDC1122,JPA35p8551}]. However, lack of controllability does not necessarily 
imply that the dynamical Lie group of the system does not act transitively on any 
kinematical equivalence class.  For instance, \emph{every} dynamical Lie group clearly 
acts transitively on the equivalence class of completely random ensembles, which consists
of the single element $\op{\rho}_R=\frac{1}{N}\op{I}_N$, where $\op{I}_N$ is the identity
matrix of dimension $N$.  

\section{Graph-connectivity criteria for controllability}

Although a characterization of the degree of controllability in terms of the dynamical
Lie algebra is useful, it can be quite time-consuming to compute the Lie algebra of a
system for large $N$.  It is therefore desirable to have criteria for controllability
that are easier to verify.   

For strongly regular Hamiltonian control systems with a control Hamiltonian of the form
\begin{equation}
  \op{H}[f(t)] = \op{H}_0 + f(t) \op{H}_1, \vspace{-1ex}
\end{equation}
it has been shown that controllability can be related to the transition graph determined
by the interaction Hamiltonian $\op{H}_1$.  
\begin{defn}
The internal system Hamiltonian $\op{H}_0$ is \emph{regular} if it has unique eigenvalues,
i.e., if each eigenvalue occurs with multiplicity one.  $\op{H}_0$ is \emph{strongly
regular} if in addition, the difference between any pair of eigenvalues is unique, i.e.,
$E_i-E_j \neq E_m-E_n$ unless $(i,j)=(m,n)$.
\end{defn}
\begin{defn}
The \emph{transition graph} of a quantum system subject to a single control field is 
obtained as follows:  Choose a Hilbert space basis $\{\ket{n}: 1\le n\le N\}$ with 
respect to which $\op{H}_0$ is diagonal and identify each state $\ket{n}$ with a vertex, 
and each non-zero element of the matrix representation of the transition Hamiltonian 
$\op{H}_1$ (with respect to this basis) with an edge.
\end{defn}
Note that the transition graph is unique \emph{only} for systems with regular Hamiltonian.
For systems with degenerate energy levels there are various choices of Hilbert space bases
that diagonalize $\op{H}_0$, which need \emph{not} correspond to equivalent transition 
graphs.  This best illustrated by an example.
\begin{exmp}
Consider a three-level $\lambda$-system, i.e., suppose we have $E_1=E_3$, $E_2>E_1$ and
\[ 
  \op{H}_0 = \left[ \begin{array}{ccc} E_1 & 0 & 0 \\
                                       0 & E_2 & 0 \\
                                       0 & 0 & E_1 
                    \end{array} \right], \quad
  \op{H}_1 = \left[ \begin{array}{ccc} 0 & d & 0 \\
                                       d & 0 & d \\
                                       0 & d & 0 
                    \end{array} \right]
\]
with respect to the canonical basis $\ket{1}$, $\ket{2}$ and $\ket{3}$.  The transition 
graph associated with $\op{H}_1$ is connected if $d\neq 0$.  

However, if we choose another basis $\ket{\psi_\pm}=(\ket{1}\pm\ket{3})/\sqrt{2}$ and 
$\ket{\psi_2}=\ket{2}$ then we obtain 
\[
  \tilde{H}_0 = \left[ \begin{array}{ccc} E_1 & 0 & 0 \\
                                       0 & E_1 & 0 \\
                                       0 & 0 & E_2 
                    \end{array} \right], \quad
  \tilde{H}_1 = \left[ \begin{array}{ccc} 0 & 0 & d \\
                                       0 & 0 & 0 \\
                                       d & 0 & 0 
                    \end{array} \right]
\]
with respect to the basis $\ket{\psi_+}$, $\ket{\psi_-}$ and $\ket{\psi_2}$.  Note that 
$\tilde{H}_0$ is diagonal but the transition graph associated with $\tilde{H}_1$ is not
connected.  Instead, we obtain a coupled two-state system $\ket{\psi_+}$, $\ket{2}$ and
a decoupled (dark) state $\ket{\psi_-}$.
\begin{center}
\setlength{\unitlength}{3947sp}%
\begingroup\makeatletter\ifx\SetFigFont\undefined%
\gdef\SetFigFont#1#2#3#4#5{%
  \reset@font\fontsize{#1}{#2pt}%
  \fontfamily{#3}\fontseries{#4}\fontshape{#5}%
  \selectfont}%
\fi\endgroup%
\begin{picture}(2791,1485)(518,-736)
\thinlines
\put(601,389){\circle{150}}
\put(601,-361){\circle{150}}
\put(1351, 14){\circle{150}}
\put(2476,389){\circle{150}}
\put(2476,-361){\circle{150}}
\put(3226, 14){\circle{150}}
\put(601,-361){\line( 2, 1){750}}
\put(1351, 14){\line(-2, 1){750}}
\put(2476,-361){\line( 0, 1){750}}
\put(526,-736){\makebox(0,0)[lb]{\smash{\SetFigFont{12}{14.4}{\rmdefault}{\mddefault}{\updefault}$\ket{1}$}}}
\put(526,614){\makebox(0,0)[lb]{\smash{\SetFigFont{12}{14.4}{\rmdefault}{\mddefault}{\updefault}$\ket{3}$}}}
\put(1276,164){\makebox(0,0)[lb]{\smash{\SetFigFont{12}{14.4}{\rmdefault}{\mddefault}{\updefault}$\ket{2}$}}}
\put(2401,539){\makebox(0,0)[lb]{\smash{\SetFigFont{12}{14.4}{\rmdefault}{\mddefault}{\updefault}$\ket{\psi_2}$}}}
\put(3151,164){\makebox(0,0)[lb]{\smash{\SetFigFont{12}{14.4}{\rmdefault}{\mddefault}{\updefault}$\ket{\psi_-}$}}}
\put(2401,-661){\makebox(0,0)[lb]{\smash{\SetFigFont{12}{14.4}{\rmdefault}{\mddefault}{\updefault}$\ket{\psi_+}$}}}
\end{picture}

\end{center}
\end{exmp}
\begin{thm} \label{thm:b} [\cite{JMP43p2051}]
If the internal system Hamiltonian $\op{H}_0$ is strongly regular and the transition 
graph determined by $\op{H}_1$ is connected then the system is controllable.
\end{thm}
A similar result was first obtained for pure-state (or wavefunction) controllability
[\cite{CP267p001}].  However, the theorem above is stronger in the sense that it 
guarantees density matrix and observable controllability, not just pure-state 
controllability.

\section{Other controllability results}

Theorem \ref{thm:b} is very useful when applicable as it is much easier to verify than
the general Lie algebraic criteria for various notions of controllability.  Note, however,
that the hypothesis of strong regularity of $\op{H}_0$ is restrictive.  Although this 
requirement can be relaxed slightly, e.g., transitions that occur with zero probability
can be ignored, etc., it essentially restricts the applicability of the result to 
systems with non-degenerate energy levels and non-degenerate transition frequencies.

Nevertheless, there are positive controllability results for systems that do not satisfy
the hypothesis of strong regularity, or even regularity of $\op{H}_0$.  For instance, 
in some cases the general Lie algebraic criteria (theorem \ref{thm:a}) can be applied 
to derive conditions on the parameters of a model system that guarantee controllability.
\begin{thm} \label{thm:c}[\cite{JPA34p1679,PRA63n063410}] 
A quantum control system with Hamiltonian $\op{H}=\op{H}_0+f(t)\op{H}_1$, where \vspace{-1ex}
\begin{eqnarray*}
      \op{H}_0 &=& \sum_{n=1}^N E_n \ket{n}\bra{n}, \\
      \op{H}_1 &=& \sum_{n=1}^{N-1} d_n (\ket{n}\bra{n+1}+\ket{n+1}\bra{n}) \vspace{-1ex}
\end{eqnarray*}
is controllable if $d_n\neq 0$ for $1\le n\le N-1$ and either
\begin{enumerate}
\item there exists $p$ such that $\omega_n\neq\omega_p$ for $n\neq p$ where
      $\omega_n=E_{n+1}-E_n$; or 
\item $\omega_n=\omega$ for all $n$ but there exists $p$ such that $v_n\neq v_p$ 
      for $n\neq p$, where $v_n \equiv 2d_n^2-d_{n-1}^2-d_{n+1}^2$.
\end{enumerate}
If $N=2p$ then $d_{p-k}^2 \neq d_{p+k}^2$ for some $k\neq 0$ is required as well. 
If in addition $\trace(\op{H}_0)\neq 0$ then the system is completely controllable.
\end{thm}
Essentially, this theorem shows that a sequentially coupled Hamiltonian quantum system
is controllable if there is a \emph{single} unique transition frequency, and this is not
the `middle' transition in case of an even-dimensional system.  If the unique transition
frequency is the middle transition and the system dimension is even, then the degree of 
controllability depends on whether the system satisfies a symplectic symmetry relation, 
in which case we have only pure-state controllability, or not, in which case we have 
general controllability [\cite{JPA35p2327}].

If there is no unique transition frequency then controllability depends on the values of
the dipole moments, but even if all transition frequencies are the same, the system is 
still controllable for most choices of the dipole moments.

Furthermore, even for systems with no non-degenerate energy levels (i.e., completely 
non-regular $\op{H}_0$) such as electronic transitions in atomic or ionic systems, 
positive controllability results have been obtained in many cases using Lie algebraic 
techniques [\cite{JPA35p4125}].

The question arises whether the hypothesis of strong regularity in theorem \ref{thm:b} 
is necessary.  Unfortunately, in spite of some encouraging results, the answer is yes.  
Regularity is \emph{not} sufficient for controllability in general.  For instance, the 
Lie algebra of a system with equally spaced but non-degenerate energy levels and uniform
transition dipole moments has been shown to be $sp(N/2)$ if $N$ even, and $so(N)$ if $N$
odd [\cite{JPA35p2327}].  Thus, the system is remains pure-state controllable if it is 
even-dimensional, but is not controllable in the odd-dimensional case.  Note that this 
result does not violate theorem \ref{thm:c} since there is no unique transition frequency
and the system does not satisfy hypothesis (b) of the theorem either.

\section{Controllability of open systems}

While there are perhaps still some interesting open problems regarding the degree of
controllability of closed quantum systems subject to Hamiltonian dynamics and open-loop 
coherent control, the results of the previous sections show that their controllability
properties are quite well understood, and that they are controllable in most cases.  
The situation is quite different for open quantum systems, i.e., systems that interact 
(incoherently) with their environment in addition to any interaction with coherent 
control fields.  Such incoherent interactions with the environment lead to a non-zero 
dissipative term $\L_D$ in the quantum Liouville equation, and inevitably result in
non-Hamiltonian dynamics.

It has been suggested that dissipative effects increase the amount of control since 
the dissipative term, being non-Hamiltonian, removes restrictions such as unitary 
evolution and tends to enlarge the dynamical Lie group of the system.  However, this
picture is only partially true.  Indeed, dissipative effects usually enlarge the Lie
\emph{algebra} of the system [\cite{qph0211027}], and states not reachable from a 
given initial state via coherent control in the non-dissipative case may become 
reachable when dissipation is added.  In fact, important applications such as optical
pumping or laser cooling rely on the combined effect of dissipation and coherent 
control fields [\cite{PRA63n013407, MTNS02p2178_4}].

However, dissipative effects generally do not increase open-loop controllability of
quantum systems.  Before we discuss why, it should be noted that it does not make 
sense to distinguish between pure and mixed state controllability for open systems, 
since dissipative effects can convert pure states into mixed states and vice versa.
Furthermore, since the kinematical equivalence classes of states that exist for 
closed systems (subject to Hamiltonian dynamics) are not preserved, the only sensible
definition of (state) controllability for open systems appears to be the following.
\begin{defn}
An open (non-Hamiltonian) quantum system is \emph{(density matrix) controllable} 
if any target state represented by a density operator $\op{\rho}_1$ can be reached
from any given initial state $\op{\rho}_0$. 
\end{defn}
It can be shown that open quantum systems are virtually \emph{never} open-loop
controllable according to this definition [\cite{qph0211027,qph0211094}].  The main
reason for this is that dissipative effects are (a) generally \emph{not} controllable
-- but rather constitute a non-Hamiltonian drift (or disturbance) term -- and always
lead to irreversible semi-group dynamics.  This means that for open systems there 
always exist states that are not accessible (via open-loop control), and the set of 
such states may be large.  

Nonetheless, non-Hamiltonian drift terms may render states not accessible from a 
given initial state by coherent control in the non-dissipative case, accessible 
via open-loop control.  Thus, it appears that the more important issue for control
of open systems is the study of the set of reachable states.  For systems with 
sufficiently small non-Hamiltonian drift terms, it may be interesting to compare 
the set of reachable states with and without drift as a function of \emph{time} 
and to consider the amount of overlap of these sets.  Since dissipative systems 
tend to asymptotically converge to a stationary state (which may depend on the 
initial state) when no control fields are applied, and coherent control operations
are unitary, it is to be expected that the set of reachable states from any initial
state will contract with time for many systems.  One question of interest in this
regard is what states are asymptotically reachable for a given open system using
open-loop control.  

\section{Closed-loop (feedback) control}

The previous sections have focused exclusively on the extent to which it is 
possible to control a quantum system (Hamiltonian or otherwise) using open-loop
coherent control.  While there are many potential applications of open-loop 
control for quantum systems, this type of control clearly has its limitations.
An alternative is to add measurements and condition the controls based on the
result of these measurements.  This type of control is generally referred to 
as closed-loop or feedback control.

Feedback control can be difficult to realize for quantum systems since feedback
requires measurements, which perturb the system and generally lead to nonlinear
dynamics.  Projective measurements give rise to discontinuous evolution and jumps
in state space.  Weak (non-projective) measurements can ameliorate the situation 
by eliminating discontinuities, but may be difficult to perform and still lead 
to non-linear dynamics.

Nevertheless, many different types of measurements and feedback are conceivable 
for quantum systems, and various schemes have recently been proposed.  A detailed
discussion of this topic is beyond the scope of a short paper such as this, but 
closed-loop control has been shown to be generally more powerful than open-loop 
control, especially for control of open system dynamics [\cite{qph0008101}].  
Feedback control using weak measurements has been proposed for tasks that are not
possible via open-loop coherent control such as continuous quantum error correction
[\cite{qph0302006}].  Furthermore, the combination of measurements and coherent 
control allows the implementation of non-unitary dynamics even for otherwise closed
systems [\cite{PRA65n010101}].  Recent results even suggest that non-dissipative 
systems that are \emph{not} open-loop controllable because the dynamical Lie group
does not act transitively on the equivalence classes of states, may be controllable 
if certain measurements are possible [\cite{qph0212006}].

\section{Conclusion}

We have presented a concise summary of recent results on open-loop controllability
of (closed) quantum systems subject to Hamiltonian dynamics, showing the importance
of the dynamical Lie algebra and related Lie group in assessing the degree of 
controllability of the system.  Negative (open-loop) controllability results of for
open quantum systems with non-Hamiltonian drift terms have been discussed, and we 
have argued for the importance of studying the reachable sets in this case.  Finally,
we have presented a brief discussion of recent results on closed-loop control of 
quantum systems, indicating the challenges of implementing feedback control, and 
demonstrating its superiority over open-loop control.

\ack
SGS would like to thank J. V. Leahy (Univ. of Oregon) and G.\ Vinnicombe (Cambridge
Univ.) for helpful discussions.

\bibliography{papers}

\begin{thebibliography}{18}
\expandafter\ifx\csname natexlab\endcsname\relax\def\natexlab#1{#1}\fi
\expandafter\ifx\csname url\endcsname\relax
  \def\url#1{{\tt #1}}\fi

\bibitem[Ahn et~al.(2003)Ahn, Wiseman, and Milburn]{qph0302006}
C.~Ahn, H.~W. Wiseman, and G.~J. Milburn.
\newblock Quantum error correction for continuously detected errors.
\newblock {\em quant-ph/0302006}, 2003.

\bibitem[Albertini and D'Alessandro(2001)]{qph0106128}
F.~Albertini and D.~D'Alessandro.
\newblock Notions of controllability for quantum-mechanical systems.
\newblock {\em quant-ph/0106128}, 2001.

\bibitem[Altafini(2002{\natexlab{a}})]{JMP43p2051}
C.~Altafini.
\newblock Controllability of quantum mechanical systems by root space
  decompositions of su(n).
\newblock {\em J. Math. Phys.}, 43\penalty0 (5):\penalty0 2051--2062,
  2002{\natexlab{a}}.

\bibitem[Altafini(2002{\natexlab{b}})]{qph0211094}
Claudio Altafini.
\newblock Controllability properties for finite dimensional quantum markovian
  master equations.
\newblock {\em quant-ph/0211094}, 2002{\natexlab{b}}.

\bibitem[Fu et~al.(2001)Fu, Schirmer, and Solomon]{JPA34p1679}
H.~Fu, S.~G. Schirmer, and A.~I. Solomon.
\newblock Complete controllability of finite-level quantum systems.
\newblock {\em J. Phys. A}, 34:\penalty0 1679, 2001.

\bibitem[Girardeau et~al.(1998)Girardeau, Schirmer, Leahy, and
  Koch]{PRA58p2684}
M.~D. Girardeau, S.~G. Schirmer, J.~V. Leahy, and R.~M. Koch.
\newblock Kinematical bounds on optimization of observables for quantum
  systems.
\newblock {\em Phys. Rev. A}, 58:\penalty0 2684, 1998.

\bibitem[Lloyd and Viola(2000)]{qph0008101}
Seth Lloyd and Lorenza Viola.
\newblock Control of open quantum system dynamics.
\newblock {\em quant-ph/0008101}, 2000.

\bibitem[Lloyd and Viola(2002)]{PRA65n010101}
Seth Lloyd and Lorenza Viola.
\newblock Engineering quantum dynamics.
\newblock {\em Phys. Rev. A}, 65:\penalty0 010101, 2002.

\bibitem[Mendes and Man'ko(2002)]{qph0212006}
R.~V. Mendes and V.~I. Man'ko.
\newblock Quantum control and the Strocci map.
\newblock {\em quant-ph/0212006}, 2002.

\bibitem[Schirmer(2001)]{PRA63n013407}
S.~G. Schirmer.
\newblock Laser cooling of internal molecular degrees of freedom for
  vibrationally hot molecules.
\newblock {\em Phys. Rev. A}, 63\penalty0 (1):\penalty0 013407, 2001.

\bibitem[Schirmer et~al.(2001)Schirmer, Fu, and Solomon]{PRA63n063410}
S.~G. Schirmer, H.~Fu, and A.~I. Solomon.
\newblock Complete controllability of quantum systems.
\newblock {\em Phys. Rev. A}, 63:\penalty0 063410, 2001.

\bibitem[Schirmer et~al.(2002{\natexlab{a}})Schirmer, Leahy, and
  Solomon]{JPA35p4125}
S.~G. Schirmer, J.~V. Leahy, and A.~I. Solomon.
\newblock Degrees of controllability for quantum systems and applications to
  atomic systems.
\newblock {\em J. Phys. A}, 35:\penalty0 4125, 2002{\natexlab{a}}.

\bibitem[Schirmer et~al.(2002{\natexlab{b}})Schirmer, Pullen, and
  Solomon]{JPA35p2327}
S.~G. Schirmer, I.~C.~H. Pullen, and A.~I. Solomon.
\newblock Identification of dynamical lie algebras for finite-level quantum
  control systems.
\newblock {\em J. Phys. A}, 35\penalty0 (9):\penalty0 2327, 2002{\natexlab{b}}.

\bibitem[Schirmer and Solomon(2001)]{IEEE40CDC1122}
S.~G. Schirmer and A.~I. Solomon.
\newblock Non-reachable target states for pure-state controllable and
  non-controllable quantum systems.
\newblock In {\em 40th IEEE CDC Proceedings} (Omnipress, Madison, WI), 
      pages 2605--2606, 2001. 
      
\bibitem[Schirmer and Solomon(2002)]{MTNS02p2178_4}
S.~G. Schirmer and A.~I. Solomon.
\newblock Quantum control of dissipative systems.
\newblock In {\em Proceedings of the MTNS 2002}, available online at
          www.nd.edu/$\sim$mtns/papers/2178{\_}4.pdf, 2002.

\bibitem[Schirmer et~al.(2002{\natexlab{c}})Schirmer, Solomon, and
  Leahy]{JPA35p8551}
S.~G. Schirmer, A.~I. Solomon, and J.~V. Leahy.
\newblock Criteria for dynamical reachability of quantum states.
\newblock {\em J. Phys. A}, 35:\penalty0 8551--8562, 2002{\natexlab{c}}.

\bibitem[Solomon and Schirmer(2002)]{qph0211027}
Allan~I. Solomon and Sonia~G. Schirmer.
\newblock Dissipative groups and the bloch ball.
\newblock {\em quant-ph/0211027}, 2002.

\bibitem[Turinici and Rabitz(2001)]{CP267p001}
G.~Turinici and H.~Rabitz.
\newblock Quantum wavefunction controllability.
\newblock {\em Chem. Phys.}, 267:\penalty0 1--9, 2001.

\end{thebibliography}
\end{document}